\documentclass{emulateapj}
\usepackage{graphicx}
\usepackage{amsmath}

\begin{document}

%\submitted{To be submitted to The Astrophysical Journal Letters}   
%\vspace{1mm}

\shortauthors{Wang J. et al.}

\shorttitle{Extended X-ray Emission in NGC 4151}


\title{Extended X-ray Emission in the HI Cavity of NGC 4151: Galaxy-scale AGN Feedback?}


\author{Junfeng Wang\altaffilmark{1}, Giuseppina Fabbiano\altaffilmark{1}, Guido Risaliti\altaffilmark{1,2}, Martin Elvis\altaffilmark{1}, Carole G. Mundell\altaffilmark{3}, Gaelle Dumas\altaffilmark{4}, Eva Schinnerer\altaffilmark{4}, and Andreas Zezas\altaffilmark{1,5}}


 \altaffiltext{1}{Harvard-Smithsonian Center for Astrophysics, 60 Garden St,
  Cambridge, MA 02138}
 \altaffiltext{2}{INAF-Arcetri Observatory, Largo E, Fermi 5, I-50125 Firenze, Italy}
 \altaffiltext{3}{Astrophysics Research Institute, Liverpool John Moores University, Birkenhead CH41 1LD, UK}
 \altaffiltext{4}{Max-Planck-Institut f$\ddot{\rm u}$r Astronomie, K$\ddot{\rm o}$nigstuhl 17, D-69117 Heidelberg, Germany}
 \altaffiltext{5}{Physics Department, University of Crete, P.O. Box 2208, GR-710 03, Heraklion, Crete, Greece}


\begin{abstract}

We present the {\em Chandra} discovery of soft diffuse X-ray emission
in NGC 4151 ($L_{0.5-2keV}\sim 10^{39}$ erg s$^{-1}$), extending
$\sim$2 kpc from the active nucleus and filling in the cavity of the
HI material.  The best fit to the X-ray spectrum requires either a
$kT\sim 0.25$ keV thermal plasma or a photoionized component.  In the
thermal scenario, hot gas heated by the nuclear outflow would be
confined by the thermal pressure of the HI gas and the dynamic
pressure of inflowing neutral material in the galactic disk.  In the
case of photoionization, the nucleus must have experienced an
Eddington limit outburst.  For both scenarios, the AGN-host
interaction in NGC 4151 must have occured relatively recently (some
$10^4$ yr ago).  This very short timescale to the last episode of high
activity phase may imply such outbursts occupy $\gtrsim$1\% of AGN
lifetime.

\end{abstract}


\keywords{X-rays: galaxies --- galaxies: Seyfert --- galaxies: jets
  --- galaxies: individual (NGC 4151)}  

\section{Introduction}

NGC 4151 \citep[$D\sim 13.3$ Mpc, $1\arcsec=65$ pc;][]{Mundell99} is
often considered the nearest and apparently brightest archetypal
Seyfert 1 galaxy (see Ulrich 2000 for a review).  Extensively observed
across the electromagnetic spectrum, it thus offers one of the best
chances of studying the interaction between an active galactic nucleus
(AGN) and the interstellar medium (ISM) in the galactic disk of its
host.  Such interaction, or ``feedback'', is recognized to play a key
role in the supermassive black hole (SMBH) and host galaxy
co-evolution \citep[e.g.,][]{Silk98}.  Direct observational
constraints are still lacking on how efficient the AGN outflows are at
depositing their energy in the host ISM
\citep[e.g.,][]{Hopkins10,Ostriker10}.

NGC 4151 has a biconical extended narrow line region (ENLR) aligned
along the Northeast--Southwest direction
\citep[P.A.$\sim$65$^{\circ}$/230$^{\circ}$;
  e.g.,][]{Evans93,Kaiser00}, which shows extended soft X-ray emission
in the central $10\arcsec$ region
\citep[e.g.,][]{Elvis83,Ogle00,Yang01,G08}. Published work on imaging
the circum-nuclear region of NGC 4151 has explored features on a few
10 pc to $\sim$1 kpc from the nucleus
\citep[e.g.,][]{Pedlar92,Mundell95,Asif98,Mundell03,Das05,Kraemer08,Wang09,SB09,SB10}.

In this Letter, we present the discovery of soft diffuse X-ray
emission extending $\gtrsim$2 kpc from the active nucleus and we
discuss its implications.

\section{Observations and Reduction}

NGC 4151 was observed by {\em Chandra} with ACIS-S (Garmire et
al. 2000) in 1/8 sub-array mode during March 27-29, 2008.  The nucleus
was placed near the aimpoint on the S3 chip.  The data were
reprocessed and analyzed following the standard
procedures\footnote{\url{http://cxc.harvard.edu/ciao/threads/createL2/}}
using CIAO (Version 4.2) and CALDB (Version 4.2.0). After removing
brief times of high background count rates, the cleaned data have
total good exposure times of 116 ks and 63 ks, in ObsID 9217 and ObsID
9218 respectively.

The two ACIS observations of NGC 4151 were then merged to create a
single event file.  The ACIS readout streaks along P.A.=
174$^{\circ}$/354$^{\circ}$ were removed with CIAO tool {\it
  acisreadcorr}\footnote{\url{http://cxc.harvard.edu/ciao/ahelp/acisreadcorr.html}}.
Point source detection was done with {\it wavdetect\/}
\citep{Freeman02} and 24 sources were removed from the images.  We
extracted the soft-band (0.3--1 keV) image and applied adaptive
smoothing using CIAO tool {\it csmooth}, with a minimum significance
3$\sigma$. The same smoothing kernel was applied to the exposure map,
which was then divided to get the exposure corrected image.

\section{Results}
\subsection{Multiwavelength Morphology}

\begin{figure}
\centering
    \includegraphics[height=.35\textheight]{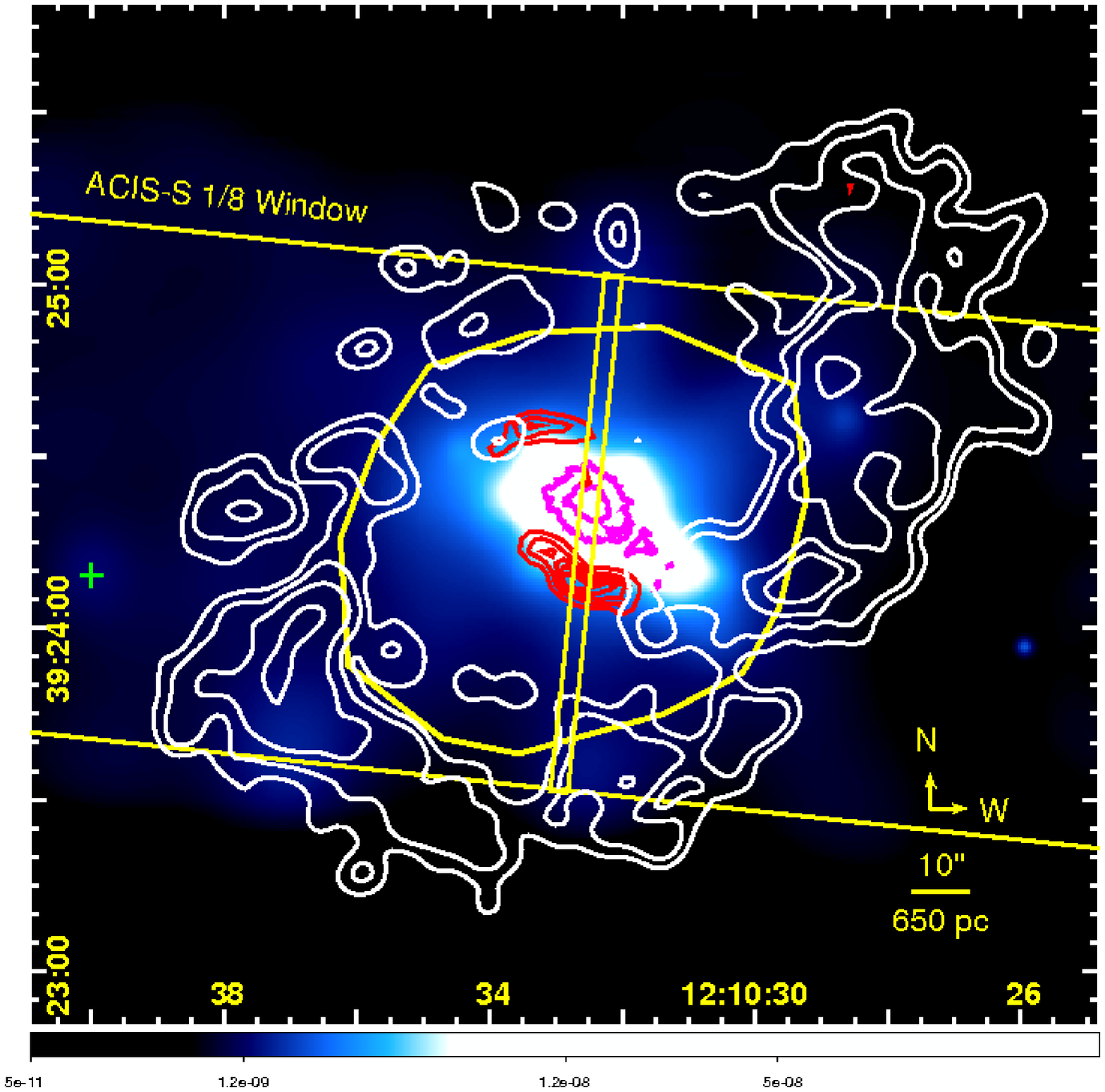}
    \includegraphics[height=.3\textheight]{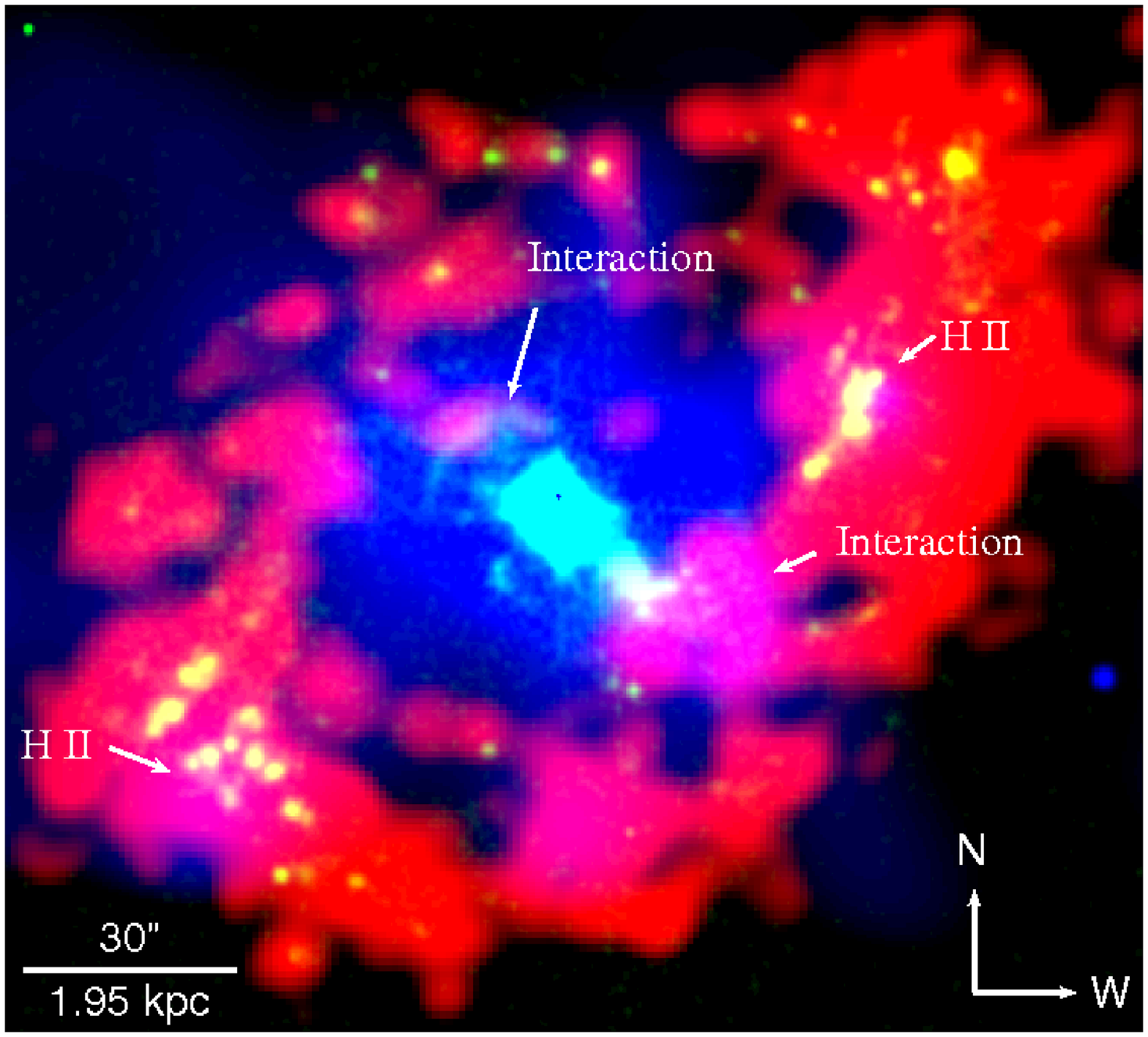}
%  \vspace{-0.8pc}
\caption{(a) Exposure corrected, adaptively smoothed ACIS image
  (0.3--1 keV) of NGC 4151, overlaid with the HI emission in contours.
  Yellow polygon outlines the spectral extraction region. Magenta
  contours indicate the spatial extent of the optical ENLR
  \citep{Kaiser00}.  Red contours represent the $^{12}$CO line
  emission \citep{Dumas10}.  A faint X-ray enhancement at
  $r=80\arcsec$ is marked with a cross.  (b) Three-color composite
  image of the central $3\arcmin\times 3\arcmin$ region. Red
  represents the HI $\lambda$21 cm emission (Mundell \& Shone 1999),
  green the continuum-subtracted H$\alpha$ emission \citep{Knapen04},
  and blue the soft X-ray emission.  Magenta (X-ray+HI) represents
  X-ray emission associated with the outflow/ISM interaction region
  and HII regions. Note that green (H$\alpha$) mostly appears as
  yellow (H$\alpha$+HI) and cyan (H$\alpha$+X-ray), which highlights
  the HII regions and the ionized bicone, respectively.} \label{fig1}
\end{figure}

Figure~\ref{fig1}a shows the resulting soft X-ray image
($3\arcmin\times 3\arcmin$, $\sim$12 kpc on a side).  Note that the
non-detection of X-ray emission towards the northernmost and
southernmost part of the image is artificial, because this area is out
of the ACIS-S field of view (FOV) for our observation
(Figure~\ref{fig1}a).  Figure~\ref{fig1}b presents a composite image
of the same region in NGC 4151, consisting of soft X-ray (0.3--1 keV),
exposure corrected {\em Chandra} ACIS image (blue), with the HI
$\lambda$21 cm map (red) and a continuum-subtracted H$\alpha$ image
\citep[green;][]{Knapen04}.

The kpc-scale HI distribution \citep{MS99} appears as a ring with
brightest emission in the NW and SE (white contours in Figure~1a).
The VLA fluxes \citep{Mundell99} agree well with lower resolution
single dish measurements \citep{Pedlar92}, supporting that the
apparent central cavity is truly devoid of HI.  Towards the nuclear
region, only localized HI absorption toward the radio jet is found
\citep{Mundell95,Mundell99}.  Inside this oval HI distribution, CO
line emission \citep{Dumas10} is prominent in two lanes $\sim$1 kpc
north and south of the nucleus (red contours in Figure~1a), and
coincident with the circum-nuclear dust ring \citep{Asif98}.

The ionized gas traced by H$\alpha$ is mainly located in the
$\sim$20\arcsec-long biconical ENLR (cyan=H$\alpha$+X-ray in
Figure~\ref{fig1}b) along the NE-SW direction centered on the nucleus,
which closely follows the high excitation emission line gas (e.g.,
[OIII]5007, Kaiser et al. 2000; magenta contours in
Figure~\ref{fig1}a) photoionized by the AGN radiation.  There is also
H$\alpha$ emission associated with a string of HII regions located
along the NW and SE edges of the large-scale stellar bar
\citep[yellow=H$\alpha$+HI;][]{Perez90}.

The soft X-ray emission is brightest in the nuclear region and central
$15\arcsec$, extending along the same direction as the ENLR bicone
(P.A.$\sim$65$^{\circ}$/230$^{\circ}$; appearing as the cyan rectangle
in Figure~\ref{fig1}b).  This emission and the associated
morphological features will be discussed in detail in a companion
paper (Wang et al., in preparation).

Two X-ray clumps are present at the terminals of the bicone (marked as
``interaction'' in Figure~\ref{fig1}b), where the outflows appear to
encounter dense materials (CO gas lane in the NE and HI clump in the
SW).  In addition, there are two regions of X-ray enhancements in the
HI ring (marked as ``HII'' in Figure~\ref{fig1}b), spatially
coincident with known HII regions \citep{Perez90}.  There are also
some anti-correlations in the spatial distribution of X-ray emission
relative to the HI material due to obscuration of the soft X-rays.
Beyond the HI material ($\sim$60$\arcsec$ from the nucleus), the X-ray
emission becomes weak to the east and absent to the west.

Here we focus on the presence of the previously unknown, faint soft
X-ray emission, which is seen extending beyond a radius of $30\arcsec$
($\sim$2 kpc), filling the cavity in the HI distribution.

\subsection{Characterizing the Soft X-ray Emission}\label{char}

We extracted X-ray data from an elliptical region centered at the
nucleus (yellow polygon in Figure~\ref{fig1}a), which covers the HI
cavity and excludes the inner brighter emission ($r\lesssim
15\arcsec$) associated with the nucleus and the optical outflow. We
measured the background level to be $0.050\pm0.006$ counts per ACIS
pixel, selecting alternative background regions in the image
$2\arcmin-3\arcmin$ away from NGC 4151.  We find that the extended
emission (0.3--1 keV) in the cavity region is significant (1562 counts
vs. 728 counts expected from background) at 11$\sigma$ ($F_{0.5-2
  keV}=3.2\times 10^{-14}$ erg s$^{-1}$ cm$^{-2}$, $L_{0.5-2 keV}\sim
10^{39}$ erg s$^{-1}$).  Moreover, we performed ChaRT\footnote{See
  \url{http://cxc.harvard.edu/chart/}} and MARX\footnote{Version 4.3;
  See \url{http://space.mit.edu/cxc/marx/}} simulations of the strong
nuclear source and the resolved extended emission using {\em Chandra}
ACIS spectra\citep{Yang01,Wang10}, which demonstrate that the point
spread function (PSF) wings of the nuclear emission and the extended
emission can only contribute 157$\pm$12 counts and 42$\pm$6 counts in
the 0.3--1 keV band in the extraction region, respectively.

We extracted the ACIS spectrum of the soft X-ray emission in the same
region.  The background subtracted X-ray spectrum is poorly fitted
with an absorbed power-law model that is consistent with the nuclear
spectrum \citep[$\Gamma=1.68$;][]{Wang10}, showing significant line
emission residuals in the 0.3--1 keV range.  Moreover, the spectrum
can be neither fitted by a single photoionized emission model
($photemis$; see XSTAR\footnote{Version 2.2; available at
  \url{http://heasarc.nasa.gov/docs/software/xstar/xstar.html}}) nor a
thermal plasma \citep[APEC;][]{Smith01} model ($\chi^2/dof=182/83$ and
$178/83$, respectively).

The fit is much improved when a power-law component with a
photoionized emission component is used ($\log \xi=1.7\pm 0.2$), where
ionization parameter $\xi \equiv L/nR^2$ \citep{KM82}.  The fit is
slightly improved when a thermal plasma (APEC; Smith et al. 2001)
component ($kT=0.25_{-0.03}^{+0.04}$ keV) is adopted.  In both
two-component fits, the power-law component contributes $\sim$50\% of
the 0.3--2 keV flux.  The results are summarized in Table~1.

\begin{figure}
\centering
    \includegraphics[height=.3\textheight,angle=-90]{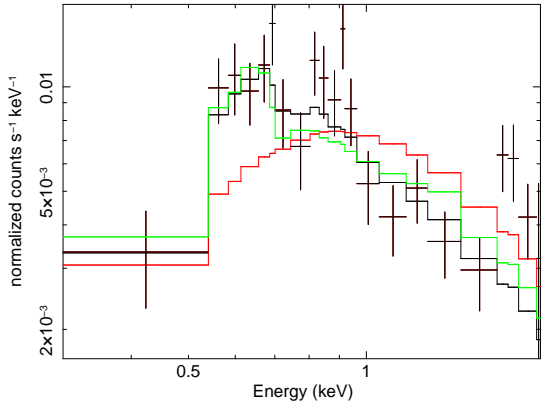}
%  \vspace{-0.8pc}
\caption{Spectral fits to NGC 4151's extended emission (0.3--2 keV),
  using a power-law component (red), a thermal plasma model (black),
  and a photoionization model (green).  The absorption is fixed at the
  line-of-sight Galactic column $N_H=2\times 10^{20}$
  cm$^{-2}$.}\label{fig2}
\end{figure}

The surface brightness profile of the extended emission is shown in
Figure~\ref{fig3}, using $\Delta r=5\arcsec$ concentric annuli
starting at $r=15\arcsec$ from the nucleus.  The small excess at
$r\sim 160$ pixel ($80\arcsec$) is due to a faint extended ($d\sim
6\arcsec$) X-ray enhancement ($L_x\sim 4\times 10^{37}$ ergs s$^{-1}$
assuming the same $kT\sim 0.25$ keV) centered at
$\alpha$=12$^h$10$^m$39.$^s$9,
$\delta$=+39$^{\circ}$24$^{\prime}$10$^{\prime\prime}$
(P.A.$\sim$120$^{\circ}$).  It appears to be spatially coincident with
the faint diffuse HI gas seen around the outer ends of the bar
connecting to the spiral arms (Pedlar et al. 1992).

\begin{deluxetable*}{lccc}
\tabletypesize{\footnotesize} \tablecaption{Spectral Fit of the NGC 4151 Extended Emission\label{tab1}} \tablewidth{0pt}
\tablehead{\colhead{Model\tablenotemark{a}} & \colhead{$\chi^2/dof$} & \colhead{Parameter} & \colhead{$F_{0.5-2 keV}$\tablenotemark{b}}}
\startdata
Electron Scattering ($pow$) & 104/83 & $\Gamma=1.68$ & 3.0 \\
Photoionization ($photemis+pow$) & 78/82 & $\log \xi=1.7\pm 0.2$ & 3.2 \\
Thermal Plasma ($apec+pow$) & 73/82 & $kT=0.25_{-0.03}^{+0.04}$ & 3.2\\
\enddata

\tablenotetext{a}{A fixed line-of-sight $N_H=2\times 10^{22}$ cm$^{-2}$ (Murphy et al. 1996) is adopted for all fits using the 0.3--2 keV spectrum.}
\tablenotetext{b}{In unit of $10^{-14}$ erg s$^{-1}$ cm$^{-2}$.}
\end{deluxetable*}

\section{Discussion}

\subsection{Nature of the Soft X-ray Emission}\label{nature}

\begin{figure}
\centering
    \includegraphics[height=.28\textheight]{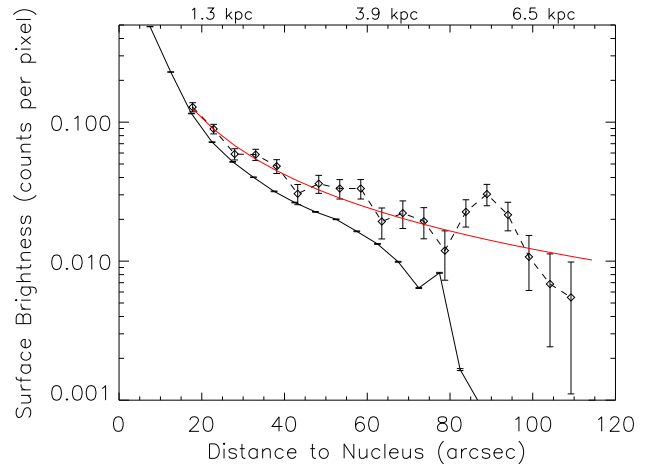}
%  \vspace{-0.8pc}
\caption{NGC 4151 surface brightness profile (background subtracted;
  dashed line) derived from ACIS 0.3--1 keV image.  The red line shows
  the best fit with a 1-d $\beta$ model.  The solid black line
  represents the K-band surface brightness profile from Knapen et
  al. (2003), resembling the integrated distribution of star light in
  NGC 4151's bulge.}\label{fig3}
\end{figure}

We consider four hypotheses for the extended soft X-rays on spatial
scales of $\sim 10^4$ light year:

{\em 1. Unresolved Point Sources}--Active star formation in the region
of the HI cavity is at a low level, as indicated by the lack of
H$\alpha$ emission and weak PAH emission \citep{Asif05}.  Adopting the
empirical measured $L_x/M_{\ast}\sim 8.2 \times 10^{27}$ erg s$^{-1}$
M$_{\odot}^{-1}$ (Revnivtsev et al. 2008) and a bulge mass of
$M_{\ast}\sim 10^9M_{\odot}$ for NGC 4151 \citep{Wandel02}, the
expected combined contribution of the unresolved old stellar
population (low mass X-ray binaries and cataclysmic variables) is
$L_{0.5-2 keV}\sim 10^{37}$ erg s$^{-1}$, two orders of magnitude
lower than the detected soft emission ($L_{0.5-2 keV}\sim 10^{39}$ erg
s$^{-1}$).  Moreover, the $K$-band starlight profile \citep{Knapen03}
decreases faster than the X-ray emission profile (Figure~\ref{fig3}).
Thus we conclude that the contribution of stellar sources is
negligible.

{\em 2. Electron Scattered Nuclear Emission}--The hot plasma around
the nucleus could electron scatter a fraction of the nuclear flux at a
larger radii (e.g., NGC 1068, Elvis et al. 1990).  Assuming a solid
angle $\Omega=4\pi$ for a scattering medium, the observed
$L_x/L_{x,nuc}$ implies an electron scattering optical depth of
$\tau_{es}=0.01$ and a mean column density of $N_H\sim 10^{22}$
cm$^{-2}$.  Adopting 200 pc for the depth, the typical galactic disk
scale height \citep[e.g.,][]{Padoan01}, this corresponds to a volume
density of $n_e\sim 15$ cm$^{-3}$.  This is $\sim$100 times higher
than $n_e$ inferred for the thermal plasma and so thermal emission
(which scales as $n_e^2$) will dominate the emission unless it is much
cooler than $10^6$~K.  The scattering medium must also be highly
ionized, requiring $T\gtrsim 3\times 10^6$ K if this is achieved
thermally.  For the electron-scattering model to be self-consistent, a
photoionized medium (eliminating $T$ discrepancy) or a historic
outburst ($\sim 10^4$ yr ago) during which the nucleus was 100 times
brighter than currently observed (eliminating $n_e$ discrepancy) is
needed.

However, the poor fit with the nuclear power-law model (\S~\ref{char})
does not support this scenario.  Moreover, the required power-law
component in either two-component fits ($L_{0.5-2 keV}\sim 4\times
10^{38}$ erg s$^{-1}$) is comparable to the expected total
contribution ($L_{0.5-2 keV}\sim 2.5\times 10^{38}$ erg s$^{-1}$) from
PSF scattering and unresolved point sources, which allows little
contribution from the electron scattered component.  We conclude that
an electron scattered component is not important for the extended
emission.
 
{\em 3. Photoionized Gas}--The faint extended emission in NGC 4151
could be due to gas photoionized by a more luminous AGN in the past.
Such a ``quasar-relic'' scenario was proposed for NGC 5252, which has
a spectacular ionization cone extending $\sim$10 kpc from the nucleus
(Dadina et al. 2010).  Observable signatures of such ``afterglow''
from an outburst in a radio quiet quasar ($L\sim 10^{46}$ erg
s$^{-1}$) have been studied by \citet{Wang05}. In this context, NGC
4151 belongs to the regime of the lower luminosity sources, for which
\citet{Wang05} suggest an X-ray spectrum dominated by Ly$\alpha$ and
Ly$\beta$ lines of NeX and OVIII.  Under this assumption, the $0.3-1$
keV emission is dominated by line emission blended at ACIS' spectral
resolution, which in turn implies an ionization parameter of $\log \xi
\sim 1.6-2.1$\citep{KM82,Yang01} consistent with $\log\xi=1.7$
measured in XSTAR.  Adopting an electron density $n=2$ cm$^{-3}$
(equal to the HI density) and $R=3$ kpc, the required ionizing
luminosity of the nucleus is $L_{ion} \sim 6\times 10^{45}$ erg
s$^{-1}\sim L_{Edd}$ ($M_{BH}=4\times 10^7M_{\odot}$; Bentz et
al. 2006).  Based on detailed photoionization modeling of the extended
optical emission of NGC 4151 \citep{SK93}, the photoionizing continuum
could be anisotropic and the ionizing flux towards the ENLR may be
$\sim$10 times higher than that in the direction of the earth.  This
is still insufficient to power the more extended X-ray emission.
Therefore an Eddington-limit outburst of NGC 4151 $\sim 10^4$ yr in
the past is required.

The same photoionized medium would produce [OIII] emission.  If we
adopt an observed [OIII]/soft X-ray ratio of $\sim$10 for $\log
\xi\sim 1.7$ photoionized gas (Bianchi et al. 2006), the expected
        [OIII] surface brightness is $\sim 10^{-16}$ erg s$^{-1}$
        cm$^{-2}$ arcsec$^{-2}$, fainter than the sensitivity limit of
        existing [OIII] images \citep[e.g., $F_{lim}\sim
          10^{-15}-10^{-16}$ erg s$^{-1}$ cm$^{-2}$
          arcsec$^{-2}$;][]{Perez90,Evans93,Kaiser00}. Thus a
        photoionized origin is still open to direct testing.

{\em 4. Confined Hot Gas}--We model the background subtracted surface
brightness profile with a 1-d $\beta$ model provided in {\tt
  SHERPA}\footnote{\url{http://cxc.harvard.edu/sherpa/}} ($\Sigma_x
\propto [1+(r/r_0)^2]^{-3\beta+1/2}$, where $r_0$ is the core
radius). This model is often adopted in studies of the hot gas in
early type galaxies \citep[e.g.,][]{Tr86}.  The best-fit index is
$\beta=0.39^{+0.03}_{-0.02}$, implying a brightness profile that is
decreasing slower than an adiabatically expanding wind
($\Sigma_x\propto r^{-3}$), and confinement of the hot gas.  We
attempted fitting spectra extracted from two concentric annuli at
radii $\Delta r=15\arcsec-30\arcsec$ and $30\arcsec-45\arcsec$. Within
the uncertainties, the thermal fits give the same temperature
($kT=0.25\pm 0.07$ keV and $kT=0.22^{+0.05}_{-0.03}$ keV), implying
that any hot gas is approximately isothermal.

Under this assumption, we derived the pressure radial profile of the
hot gas based on the surface brightness profile, considering two cases
for the volume in which the hot gas is contained: a cyclinder or a
sphere.  For the cylinder, we assumed a depth of 200 pc.  To calculate
the volume emission density of spherically symmetric gas, we account
for the projection effect by successively subtracting outer shells to
smaller radii \citep{Kriss83}.  The electron density distribution is
then derived from the emission per unit volume, adopting the simple
approximation $\epsilon \approx 6.2\times 10^{-19}T^{-0.6}n_p^2$ erg
s$^{-1}$ cm$^{-3}$ \citep{MC77} for the emissivity of a $kT=0.25$ keV
thermal gas (\S~\ref{char}).

The pressure of the X-ray gas gradually decreases from $10^{-11}$ to
approximately $\sim 5 \times 10^{-12}$ dyne cm$^{-2}$ when reaching
the HI gas lanes.  Without confinement, the gas will inevitably
expand, cooling too much to be observed in the X-rays.  We estimate
that the thermal pressure of CO is $\sim 10^{-15}$ dyne cm$^{-2}$,
using the mass measured in \citet{Dumas10}, thus the molecular gas
present ($T\sim 10$~K) cannot provide the pressure balance.  The
estimated thermal pressure of the HI material is $\sim 2\times
10^{-12}$ dyne cm$^{-2}$, assuming $T\sim 8000$~K\footnote{This value
  is adopted from measurements in our galaxy, considering turbulence
  over the galaxy and typical disk line widths (e.g., Sellwood \&
  Balbus 1999). Note that the average brightness temperature of HI in
  our galaxy is $T\sim 100-150$~K (Clark 1965).} and a column density
$N_H=1.5\times 10^{21}$ cm$^{-2}$ \citep{Mundell99}.  This is well
matched to the pressure required to provide the confinement needed to
prevent the hot gas from expanding in the disk plane where atomic gas
is most dense.

\citet{Mundell99} showed that the isovelocity contours of the HI
emission deviate from a circularly rotating disk, and identified
kinematic evidence for the presence of shocks in inflowing gas along
the stellar bar.  Taking an average inflow velocity of v$\sim$40 km
s$^{-1}$ \citep{Mundell99} and a particle density of $n=2$ cm$^{-3}$
in the HI gas lanes, the dynamic pressure as a result of the inflowing
motion, is $\rho v^2 \sim 10^{-11}$ dyne cm$^{-2}$, where $\rho=nm_p$
is the density.  This provides additional pressure for the neutral
material to be in equilibrium with the hot gas.

\subsection{Implications on the Timescale of AGN-host Interaction}

The presence of soft X-ray emission (either thermal or photoionized)
on a $\gtrsim$2 kpc scale is interesting.  If the X-ray emission
originates from hot gas heated by the AGN outflow, it would be strong
evidence for AGN feedback to the ISM on galaxy-scales, resembling the
larger scale AGN feedback from radio bubbles interacting with the
intracluster medium seen in the Perseus cluster \citep{Fabian03}.

In the photoionized scenario, considering the light travel time from
the nucleus to the HI gas lanes ($10^4$ yr) and the photoionization
recombination timescale $t_{rec}\approx 200 (n/10^9)^{-1}
(T/10^5)^{0.7}$ s \citep[$\sim 1.5 \times 10^4$ yr under the above
  assumptions;][]{Reynolds97}, we can constrain the timescale when NGC
4151 was experiencing such an Eddington-limit outburst phase to be
$t_{Edd}< 2.5\times 10^4$ yr ago.  Otherwise the X-ray gas would no
longer emit recombination lines.

If the soft diffuse emission is due to shock heating associated with a
nuclear outflow instead, the current $L_{bol}=7.3\times 10^{43}$
(Kaspi et al. 2005) indicates $L_{bol}/L_{Edd}\sim 0.01$.  Assuming
that $\sim 5\%$ of the power has gone to create the HI cavity
\citep[e.g.,][]{Silk98,Hopkins05}, $\sim 4\times 10^4$ year of such
AGN heating is needed to produce the thermal energy content of the hot
gas ($E_{th}\sim 3\times 10^{54}$ erg).  This is much shorter than the
current cooling time of the hot gas ($\tau_c\sim 10^8$ yr).  The
efficiency at which the AGN outflow deposits its kinematic energy in
the ISM can be as low as $10^{-4}$, provided that the hot gas is
confined and the duration of the strong nuclear outflow is
$\tau_{outflow}\ll \tau_c$.

However, perpendicular to the plane, the gas could expand and cool
efficiently via adiabatic expansion unless some other confinement
exists.  Assuming an outflow velocity of v$_{outflow}\sim 10^3$ km
s$^{-1}$ \citep[typical of the NLR clouds;][]{Kaiser00} for the free
expansion perpendicular to the disk (c.f. thermal velocity dispersion
v$_{th}=\sqrt{k_BT/m_p}\sim 100$ km s$^{-1}$), the relevant timescale
is the time span in which the hot gas expands to the scale height
above disk plane $\Delta R/{\rm v}_{outflow}\sim 10^5$ yr.  This
timescale again implies that the timescale to the last AGN outflow
heating must be $<10^5$ yr, in order to replenish the X-ray gas
expanding out of the plane that is cooling rapidly.

Compared to the estimated AGN lifetime ($10^7 - 10^8$ yrs), the fact
that we observe the AGN-host interaction is intriguing given the short
timescale to the last episode of activity.  Such episodic outbursts
must occur frequently \citep{Ciotti10}, and our finding implies they
may occupy $\gtrsim$1\% of the AGN lifetime.

\section{Conclusions}

To summarize, we have discovered soft (0.3--1 keV) diffuse X-ray
emission ($L_{0.5-2 keV}\sim 10^{39}$ erg s$^{-1}$) in the central
$\sim$2 kpc of NGC 4151, which cannot be attributed to PSF scattering
or electron scattering of the nuclear emission, or the integrated
emission from unresolved faint point sources.

If the gas is of photoionized origin and represents the relic of past
AGN activities, the nucleus of NGC 4151 must have experienced an
Eddington-limited high luminosity phase.  Alternatively, AGN outflows
may have mechanically heated the gas to X-ray emitting temperature
($kT=0.25$ keV). This hot gas could be confined in the HI cavity by
the thermal pressure of the HI gas and dynamic pressure of infalling
gas along the large-scale stellar bar.

For both scenarios, the AGN-host interaction must have occured
relatively recently.  For the AGN outflow heating to work, the deposit
of mechanical energy must have happened $\lesssim 10^5$ yr ago to
replenish the hot gas, which is expanding out of the plane unless
prevented by other confining mechanism.  Whereas for photoionized gas
from a past AGN outburst, the timescale to the highly active phase
must be $\lesssim 2.5\times 10^4$ yr ago.  This short timescale to the
last episode of high activity phase may imply such outbursts occupy
$\gtrsim$1\% of the AGN lifetime.

\acknowledgments

We thank the anonymous referee for helpful comments.  This work is
partially supported from NASA grant GO8-9101X and NASA Contract
NAS8-39073 (CXC).  GD acknowledges support from DFG grants SCH 536/4-1
and SCH 536/4-2 as part of the SPP 1177.
%
%\clearpage
%\newpage
%

%\clearpage

\end{document}